# The largely unconstrained multiphase nature of outflows in AGN host galaxies

Observations and simulations show that outflows in active galactic nuclei (AGN) contain gas in different phases. To understand their true impact on galaxy evolution, we advocate consistent and unbiased investigation of these multiphase winds in large AGN samples.

Claudia Cicone, Marcella Brusa, Cristina Ramos Almeida, Giovanni Cresci, Bernd Husemann and Vincenzo Mainieri

The interstellar medium (ISM) of galaxies includes gas in various phases, spanning from the cold and dense molecular clouds to the very hot highly ionised medium. Analogously, the outflows developing in AGN host galaxies have a multiphase nature, as revealed by observations and expected from simulations. The average properties of the different outflow phases that are discussed in this Comment are summarised in Table 1.

Understanding the complex interplay between the different gas phases involved in galactic winds (see Figure 1) can provide key information about the AGN feedback mechanism. For example, a large fraction of the gas mass in outflows may reside in molecular ($H_2$) clouds. Indeed, molecular line studies have delivered very high $H_2$ mass-loss rates of several 100 $M_{Sun}$/yr (refs [1,2]), which may therefore have a large impact on the star formation in the host galaxy. However, because detecting $H_2$ outflows is particularly challenging, existing analyses are inhomogeneous and suffer from several selection biases. In particular, most current investigations have relied on observations of CO or OH lines in starburst galaxies characterised by high $H_2$ gas fractions. As a result, molecular outflows that are less massive or slower with respect to other gas phases may have escaped identification. Even with the most sensitive mm/sub-mm interferometers (ALMA, NOEMA), large integrations are needed to detect the faint signature of outflows through CO line emission, which is usually >10 times fainter than that from non-outflowing gas. Furthermore, it is particularly crucial to recover the most extended and diffuse component of molecular outflows, which may carry the bulk of their mass and may be missed by interferometric observations that do not include short antenna separations. Another limitation is the uncertainty on the outflow mass estimates based on CO or OH observations in the absence of additional $H_2$ tracers. This uncertainty can reach one order of magnitude, and possibly be even higher when only high excitation transitions are available (such as for high-redshift studies based on high-J CO lines). All these technical challenges, and the large observing time needed to tackle them, have so far hindered the possibility of performing statistically significant studies of the molecular component of multiphase outflows.

Large observational efforts have been undertaken to investigate the ionised phase of galactic outflows, usually traced by broad components of rest-frame optical emission lines. Current studies include tens of thousands of low-z AGN (mostly from the Sload Digital Sky Survey [3]) and a few hundreds of high redshift AGN up to z~3 (refs [4,5]). It has been suggested that in very luminous AGN (AGN bolometric luminosity, $L_{Bol}$>$10^{46}$ erg/s) ionised outflows are ubiquitous [3], and that their mass-loss rates of ≥100 $M_{Sun}$/yr are comparable to those observed in the molecular phase [6]. However, the challenge for these studies is to obtain reliable constraints on the energetics of the ionised outflows, due to a combination of metallicity, extinction, and excitation effects (see Perspective by Harrison et al. [7] in this focus issue). For example, the uncertainties on the electron densities alone can lead to errors in ionised outflow masses that may exceed by far one order of magnitude. In addition, determining the ionised outflow geometry and spatial extent from ground-based observations is made more difficult by beam smearing effects, especially at high redshift and in absence of high sensitivity and spatially resolved data. Finally, ionised outflow studies are not immune from selection biases and have been mostly conducted on optically ([OIII]) selected samples. Instead, to fully characterise the incidence of this phenomenon in AGN, large samples spanning a wide range of intrinsic population properties (such as $L_{Bol}$) are needed.

Multiphase tracers allow us not only to probe galactic outflows in their full extent, that is, from the nuclear (< 1pc) to the largest scales (>10 kpc), but also to have a comprehensive view of their driving mechanism. Theoretical models of AGN feedback ascribe an important diagnostic role to very hot ultra-fast outflows, detectable through absorption lines associated with highly ionised Iron atoms (FeXXV and FeXXVI) observable in the hard X-ray band at energies E>6 keV (ref [8]). These features are produced by the innermost gas close to the accretion disc (<1 pc) that first feels the pressure from the AGN radiation field. Estimating the energetics of these nuclear winds and comparing it with that of larger-scale outflows provides crucial constraints on the amount of energy that is transferred to the kpc scales. Such a comparison has so far been attempted only in a few sources [9], and without fully taking into account the effects of AGN variability. Indeed, the latter may significantly complicate the relation between the outflow energetics and the AGN activity registered at a given epoch. Besides informing us about the nuclear winds, the hard X-ray band may enclose key information on the effects of AGN on the kpc-scale medium. Shocks generated by AGN-driven outflows are expected to produce a hot (T=$10^7$ K) and extended (>1 kpc) gas component [10], which would be missed by optical observations but could be revealed through imaging at 1-10 keV. However, the resolution (both spectral and spatial) and sensitivity offered by current X-ray satellites is inadequate for outflow science at kpc scales, especially at high redshift, and observations are already pushing the technical limitations of these facilities. Upcoming X-ray satellites such as Athena, the X-ray Astronomy Recovery Mission (XARM), and the Chandra Successor Mission (CSM) will offer a unique observational test for models of AGN feedback.

Detailed multiphase outflow studies have been undertaken only in a few AGN. For example, observational efforts focusing on radio galaxies have clearly demonstrated the multiphase nature of outflows when associated with the expansion of a radio jet into the surrounding medium. In this class of objects, the presence of a radio source enables the detection of the neutral atomic phase of outflows through high-velocity HI 21cm line components revealed in absorption against the continuum emission. However, the distribution of the absorption features depends strongly on that of the underlying radio continuum. In the local Seyfert Type 2 galaxy IC5063, the ionised, neutral atomic, and molecular phases of the outflow show a remarkable similarity in their kinematics and spatial extent [11], hence they are consistent with being different faces of the same feedback event. In a few other AGN where a multiphase characterisation of the outflow has been attempted, such as the nearby ultra-luminous infrared galaxy and quasar Mrk231, the picture appears significantly more complex than that emerging for IC5063, and it is not yet clear how the different outflow phases relate to each other [12].

Besides improving the statistics of current investigations and combining multi-wavelength data, exploring alternative tracers may help compensate for the technical limitations of the methods that have so far dominated outflow studies. For example, the [CII]158$\mu$m emission line is the main coolant of the ISM over a wide range of physical conditions, being at the same time little affected by extinction. At z>4 the [CII] transition is redshifted to sub-mm/mm wavelengths that have a good atmospheric transmission, making it a very promising tracer of neutral atomic and molecular outflows in the early Universe. Furthermore, the availability of sensitive integral field instruments such as MUSE at the Very Large Telescope (VLT), offering broad wavelength coverage and large field of view, allows us to simultaneously study multiple phases of the outflows (ionised from e.g. [OIII] and neutral atomic from NaID) across the whole extent of nearby galaxies. With the advent of the James Webb Space Telescope (JWST) it will be possible to undertake the study of ionised outflows in very high redshift sources (z>4), as well as to place robust constraints on the density and temperature of the ionised gas component across a broad redshift range. Furthermore, the JWST will allow us to further explore the mid-infrared rotational and ro-vibrational transitions of $H_2$ as tracers of AGN-driven outflows. These lines reveal gas at T~100-1400 K cooling out of a hot shocked medium, and so they may be enhanced in outflow environments with respect to the average galactic ISM conditions. In terms of physical scales, the lower extinction that affects infrared wavelengths enables us to trace regions closer to the base of the winds [13]. Dust is expected to participate to the outflow if $H_2$ molecules are reformed within the high-velocity clouds (see Comment by Cresci & Maiolino [14] in this focus issue). Balmer decrement studies have suggested a higher extinction in galactic winds than in the quiescent ISM [15], and mm/sub-mm observations may be able to directly detect the emission from the cold dust component of outflows.

We argue that measurements based on a single-phase view of galactic winds are often incomplete, which may result in misleading conclusions about the physics of AGN-driven outflows. In particular, neglecting the multiphase and overall complex nature of the winds leads to wrong estimates of their extent, mass, and energetics, and therefore misinterprets their relevance for galaxy evolution. A multiphase investigation of outflows in a statistical sample is a crucial next step to undertake in the field of AGN feedback studies. Achieving this goal requires that a significant amount of observing time from the X-ray to the radio bands is invested in large systematic surveys, and we advise the community to commit to such enterprise.

*Claudia Cicone is at INAF - Osservatorio Astronomico di Brera, Via Brera 28, I-20121 Milano, Italy. Marcella Brusa is at the Department of Physics and Astronomy, University of Bologna, Via Piero Gobetti 93/2, I-40129, Bologna, Italy. Cristina Ramos Almeida is at the Instituto de Astrofísica de Canarias, Calle Vía Láctea, s/n E-38205 La Laguna, Tenerife, Spain. Giovanni Cresci is at INAF – Osservatorio Astrofisico di Arcetri, Largo Enrico Fermi 5, I-50125 Florence, Italy. Bernd Husemann is at the Max-Planck-Institute for Astronomy, Königstuhl 17, D-69117 Heidelberg, Germany. Vincenzo Mainieri is at the European Southern Observatory, Karl-Schwarzschild-Strasse 2, D-85748 Garching bei München, Germany.*

**Acknowledgements:**

We thank all the participants of the Lorentz Centre workshop 'The reality and Myths of AGN feedback' held in Leiden, 16-20 October 2017, for stimulating discussion that contributed to develop the ideas presented in this Comment. C.C. acknowledges funding from the European Union's Horizon 2020 research and innovation programme under the Marie Skłodowska-Curie grant agreement No. 664931. M.B. acknowledges support from the FP7 Career Integration Grant 'eEASy' (CIG 321913). C.R.A. acknowledges the Ramón y Cajal Program of the Spanish Ministry of Economy and Competitiveness through project RYC-2014-15779 and the Spanish Plan Nacional de Astronomía y Astrofísica under grant AYA2016-76682-C3-2-P. We thank Raffaella Morganti and Emanuele Nardini for providing the data used in Figure 1.

**Table 1**: Properties of the different outflow phases discussed in this Comment

| Outflow gas phase | Primary tracers | $<T_{gas}>$ [K] | $<n_{gas}>$ [particles (e-, HI, $H_2$) cm$^{-3}$] |
|---|---|---|---|
| **Highly ionised** | X-ray absorption lines | $10^6 - 10^7$ | $10^6 - 10^8$ |
| **Ionised** | [OIII]; H$\alpha$ | $10^3 - 10^4$ | $10^2 - 10^4$ |
| **Neutral atomic** | HI 21cm; NaID; [CII] | $10^2 - 10^3$ | $1 - 10^2$ |
| **Molecular** | CO; OH; [CII]; $H_2$ IR lines | $10 - 10^2$ | $\geq 10^3$ |

**Figure 1.** *[Image not included in this version prepared by the authors due to copyright restrictions. Links to published version are provided in the ArXiv 'Comments' field]. Artistic view and observational evidence of multiphase AGN-driven outflows.* The cartoon highlights the different phases that are involved in the outflow, inspired by observations and models of AGN feedback [10]. The outflow propagates from the central engine (<1pc; panel a), through the surrounding ISM (1pc-1kpc; panel b), out to the boundaries of the host galaxy (>10 kpc; panel c). Panels d,e,f show outflows observed in different gas phases and at different scales in three well-studied AGN: the ultra-fast accretion disc wind of highly ionised gas in PDS456 (ref [16], panel d); the neutral outflow observed in the molecular and atomic phases on scales of a few hundreds of pc in Mrk231 (refs. [12,17], panel e); the ionised wind extending over a few kpc detected in NGC1365 (ref [18], panel f).